\documentclass[%
 reprint,
 amsmath,amssymb,
 aps,
]{revtex4-2}

\usepackage{graphicx}
\usepackage{dcolumn}
\usepackage{bm}

\begin{document}


\title{Early warning signals for phase transitions in networks}

\author{A. V. Goltsev}
\email{goltsev@ua.pt}
\affiliation{Department of Physics, I3N, University of Aveiro, 3810-193 Aveiro, Portugal\\ Ioffe Physical-Technical Institute, 194021 St. Petersburg, Russia
}
\author{S. N. Dorogovtsev}%
 \email{sdorogov@ua.pt}
\affiliation{
 Department of Physics, I3N, University of Aveiro, 3810-193 Aveiro, Portugal \\ Ioffe Physical-Technical Institute, 194021 St. Petersburg, Russia
}

\date{\today}

\begin{abstract}
The percolation phase transition in complex network systems attracts much attention and has numerous applications in various research fields. 
Finite size effects smooth the transition and make it difficult to predict the critical point of appearance or disappearance of the giant connected component. For this end, we introduce the susceptibility of arbitrary random undirected and directed networks and show that a strong increase of the susceptibility is the early warning signal of approaching the transition point. Our method is based on the introduction of  `observers', which are randomly chosen nodes monitoring the local connectivity of a network. To demonstrate efficiency of the method,  we derive explicit equations determining the susceptibility and study its critical behavior near continuous and mixed-order phase transitions in uncorrelated random undirected and directed networks, networks with dependency links, and $k$-cores of networks. The universality of the critical behavior is supported by the  phenomenological Landau theory of phase transitions. 
\end{abstract}

\maketitle


\section{\label{sec:intro} Introduction}
The percolation phase transition is the fundamental critical phenomenon, which has been observed in  many real complex systems including  complex networks  \cite{stauffer2003introduction,newman2010networks,dorogovtsev2022nature}. It  attracts much attention and finds numerous applications in various research fields in mathematics, physics, biology including brain and sociology. It is well  known that a complex system, which undergoes a continuous phase transition, responses strongly near the critical point to an even weak external field conjugated to the order parameter. The classical example is a sharp increase of the zero-field magnetic susceptibility in ferromagnetic materials when approaching the Curie temperature, see, for example, in Ref. \cite{baxter1982exactly}. The importance of this universal phenomenon is that it serves as an early warning signal of a continuous phase transition from a disordered to ordered magnetic state. In the case of the percolation phase transition in networks, the zero-field susceptibility can be defined as the mean size of a cluster to which a randomly chosen node belongs, see Ref.~\cite{stauffer2003introduction}. This parameter demonstrates a sharp increase when approaching the percolation point. Another equivalent definition of the zero-field susceptibility is the probability that two randomly chosen nodes belong to the same cluster. These definitions use the equivalence of the one-state Potts model to the bond percolation problem in lattice \cite{kasteleyn1969introduction,fortuin1972random,wu1982potts,lee2004evolution}. However, to use this definition of the zero-field susceptibility, it is necessary to know evolution of the network structure (organization of finite clusters and the giant connected component) during the percolation process.  This can be time-consuming for some types of networks, for example, such as directed networks. 
In the case of bond percolation on undirected networks, it was proposed to use a `ghost'  field as a conjugated field acting on nodes. The susceptibility is the response of a network on the field \cite{aharony1980universal}. Note that the ghost field is not a real field. It represents a tag that probes clusters and must be small enough.  To introduce a `ghost' field, one adds either an extra Potts state (the `ghost') or a `ghost' node connected to every node with a small probability. 

In this paper, we consider site percolation and introduce the susceptibility of arbitrary random undirected and directed complex networks. 
In contrast to the ghost field, our method is based on the introduction of so called `observers', which are randomly chosen nodes, which monitor the local network connectivity and determine how many nodes are reachable by following the edges, starting from the observers. We define the susceptibility as the ratio of the change of the number of nodes reachable from the observers to the change of the number of observers. The fraction of the observers is a real parameter, which serves as the strength of the field conjugated to the giant connected component of a network. It takes values from zero to one. The proposed method can be used in numerical simulations of percolation and does not require prior knowledge of the cluster organization of the network.
Our theoretical analysis demonstrates that the susceptibility is completely determined by the evolution of finite clusters near the percolation point, when  the giant connected component emerges or disappears. To demonstrate efficiency of the method,  we derive explicit equations determining the susceptibility as a function of removed nodes, the strength of the conjugated field, and other control parameters in uncorrelated random undirected and directed networks and networks with dependency links, the latter were introduced in \cite{parshani2011critical}. We consider continuous and mixed-order (first-order) phase transitions in the networks. The behavior of the susceptibility near the critical percolation point is determined by the critical properties of the transitions that allows to use the behavior as an indicator of the transitions. Finally, we show that the critical behavior of the susceptibility in networks completely agrees with the critical behavior predicted by the phenomenological Landau mean-field theory of phase transitions in application to complex networks \cite{goltsev2003critical,dorogovtsev2008critical} that evidences for the universality of the critical behavior in complex networks.

\section{\label{sec:arbitrary} Susceptibility of an arbitrary network}

In this section we introduce the susceptibility of arbitrary undirected and directed random networks under a process when a node is occupied with the probability $p$ and is removed with the complementary probability $1-p$.

\subsection{\label{sec:undirected} Susceptibility of an arbitrary undirected network}
We consider a network $\hat{G}$ formed by $N$ nodes linked by undirected  edges. The nodes are numerated by the index $i$, $i=1,2,\dots N$. We chose at random a fraction $h$ of the nodes. The chosen nodes 
form a set $\hat{C}_h$ of the size $hN$. The parameter $h$ also has the meaning of the probability that a randomly chosen node belongs to the set $\hat{C}_h$. Let us consider how nodes from the set are connected to other nodes in the network $\hat{G}$. Namely, for each $i \in \hat{C}_h$ we find a set of nodes in $\hat{G}$ which are reachable by following edges, starting from the node $i\in \hat{C}_h$. This can be done by use of the adjacency matrix of $\hat{G}$. We call this set as $\hat{T}(i)$. Note that $i \in \hat{T}(i)$. Equivalently, $\hat{T}(i)$ is the set of nodes, which can reach the node $i$ by following edges in $\hat{G}$. Uniting the sets $\hat{T}(i)$, we get a set of nodes, which can reach at least one of the nodes in $\hat{C}_h$:
\begin{equation}
 \hat{T}_h \equiv \bigcup_{i \in \hat{C}_h} \hat{T}(i). 
 \label{1sa}
\end{equation}
Note that that $\hat{C}_h \in \hat{T}_h$. We denote the number of nodes in $\hat{T}_h$ as $V(\hat{T}_h)$ and 
the fraction  of the nodes as 
\begin{equation}
S(h) \equiv  \frac{1}{N} V(\hat{T}_h).     
 \label{0sa}
\end{equation}
We call the nodes belonging to $\hat{C}_h$ as `observers' and the set $\hat{C}_h$ as `the field of observers' of the percolation process in the considered network. It is important to outline that the observes are ordinary nodes participating in the percolation process in $\hat{G}$. They only have an additional function to monitor the percolation process and give information about a change of the local connectivity during the process.
In this paper, we show that studying the change of  the local connectivity of the observers to other nodes during the percolation process allows to find the early warning signal for the percolation transition in complex networks. It also gives information about the kind of continuous and mixed-order phase transitions. 
The fraction $h$ of the observers plays a role of the strength of the field conjugated to the order parameter, which is the giant connected component. The larger the $h$ the larger is the fraction $S(h)$  of the nodes, which are reachable starting from the observers belonging to $\hat{C}_h$. 
We define the susceptibility of the network as follows, 
\begin{equation}
 \chi(h) \equiv  \frac{dS(h)}{ dh}.  
 \label{2sa}
\end{equation}
Thus, the susceptibility is the ratio of the change of the number of nodes reachable from the observers to the change of the number of observers. 

The method, which is formulated by Eqs.~(\ref{1sa})-(\ref{2sa}),  can be applied for finding the susceptibility of an arbitrary undirected network. In numerical simulations, it is convenient to introduce a set of nested subsets of the observers, $\hat{C}_{h_n}$: $ \hat{C}_{h_1}\subset \hat{C}_{h_2} \subset  \dots \hat{C}_{h_{max}}$, where $h_n= n \Delta$, $n=1, ... n_{max}$, and $\Delta \ll 1$. Then, $\chi(h_n) = (S(h_{n+1})-S(h_{n}))/\Delta$. If $N$ is large but finite, then it is necessary to average $S(h)$ over different realizations of the set $\hat{C}_h$. Our method assumes that the number of observers must be sufficiently large, that is, $hN \gg 1$ though $h \ll 1$.  

Let us prove that $\chi(h=0)$ equals the zero-field susceptibility defined in \cite{stauffer2003introduction} as the mean size of a cluster to which a randomly chosen node belongs. In a general case, a network consists of a set of finite clusters $\hat{s}_{\alpha}$ of interconnected nodes and the giant connected component. The size of a finite cluster $\alpha$ is $s_{\alpha}$. Each cluster $\hat{s}_{\alpha}$ can contain $n_{\alpha}$ nodes belonging to the set $\hat{C}_h$. The total number of nodes, which can reach at least one observer, is the total size of the clusters $s_{\alpha}$, which contain at least one observer  observers. Therefore, the fraction $S(h)$ we are looking for is
\begin{eqnarray}
S(h) &=&  \frac{1}{N} \sum_{\alpha} s_{\alpha} [1-(1-h)^{s_{\alpha}}]
\\
&=& \frac{1}{N} \sum_{s} N(s) s [1-(1-h)^s].
\label{4sa}
\end{eqnarray}
Here, the value $1{-}(1{-}h)^{s_{\alpha}}$ is  the probability that the cluster of size $s_{\alpha}$ contains at least one observer. It also has a meaning of the probability that the cluster $\alpha$ contributes to $S(h)$. $N(s)$ is the number of clusters of size $s$ in the considered network. Note that the summation over $s$ includes both finite clusters and the  giant connected component. The normalization condition is  
\begin{equation}
\frac{1}{N} \sum_{s} N(s) s=1.
\label{5sa}
\end{equation}
In the case $h \ll 1$, the Eq.~(\ref{4sa}) takes a form
\begin{equation}
S(h) \simeq \frac{1}{N} \sum_{s} N(s) s [1-e^{-hs}].
\label{5sa}
\end{equation}
This is the basic equation of the `ghost' field approach \cite{aharony1980universal}. 
Within our approach, the field strength $h$ is a real parameter having meaning of the fraction of the observers. This parameter has also a meaning of the probability that a randomly chosen node is the observer. This parameter does not affect percolation at any values in the interval $0<h \leq 1$. In contrast to our approach, the ghost field is not a real parameter and must be small enough not to affect percolation. 

Using Eq.~(\ref{4sa}), we find the susceptibility Eq.~(\ref{2sa}), 
\begin{equation}
\chi(h)= \frac{1}{N} \sum_{s} N(s) s^2 (1-h)^{s-1}.
\label{6sa}
\end{equation}
The contribution of the giant connected component equals $S^2 (1-h)^{S-1}/N$, where $S$ is its size and $S/N$ is the fraction of nodes belonging to it. This contribution tends to zero in the limit $N \rightarrow \infty$, because  
\begin{equation}
 S^2 (1-h)^{S-1}/N \simeq (S/N)^2 \exp(\ln N - S|\ln(1-h)|) \rightarrow 0.
 \label{6sal}
\end{equation}
We conclude that only finite clusters contribute to $\chi(h)$, 
\begin{equation}
\chi(h)= \frac{1}{N} \sum_{s_f} N(s_f) s_{f}^2 (1-h)^{s_f-1},
\label{6saf}
\end{equation}
where the summation is over  finite clusters $s_f$. Therefore, the susceptibility $\chi(h)$ is determined by the distribution function $N(s)$ of the finite clusters over size. The zero-field susceptibility is
\begin{equation}
\chi(0)= \frac{1}{N} \sum_{s_f} N(s_f) s_{f}^2,
\label{7sa}
\end{equation}
This equation agrees with  the definition of the zero-field susceptibility in \cite{stauffer2003introduction,aharony1980universal}.

Let us consider a site percolation in an uncorrelated
random tree-like network, in which nodes are removed at
random with the probability $1-p$. The fraction $p$ of nodes
remains in the network.
The probability, that a randomly chosen node belongs to a cluster of size $s$, is  $\Pi(s)=sN(s)/N$. It  has the following asymptotic behavior:  
\begin{equation}
\Pi(s) = A s^{-\alpha} \exp(-\frac{s}{s^{\ast}}),
\label{8sa}
\end{equation}
where the parameter $1/s^{\ast} =0$ at the percolation threshold $p=p_c$ and $\alpha$ is a characteristic index \cite{newman2001random}. In a general case, at $p$ near $p_c$, we have $1/s^{\ast} (p) \sim |p-p_c|^{\nu}$ with an exponent $\nu$.  Substituting this asymptotics into Eq.~(\ref{6saf}), at $h \ll 1$ we find the asymptotic behavior of $\chi(h)$,
\begin{equation}
\chi(h) \sim 1/[h+1/s^{\ast}]^{2-\alpha}.
\label{9sa}
\end{equation}
At the percolation point we obtain 
\begin{equation}
\chi(h) \sim \frac{1}{h^{2-\alpha}}.
\label{10sa}
\end{equation}
In the limit $h \rightarrow 0$, at $\alpha < 2$ the susceptibility $\chi(h)$ diverges. In the case of random uncorrelated networks  with a finite second moment of the degree distribution, the  exponent $\alpha $ is 3/2, see in Ref.~\cite{newman2001random}, as a result, $\chi(h) \sim 1/\sqrt h$ in agreement with the ghost field approach \cite{aharony1980universal,cirigliano2024scaling}. An analytical method to calculate $\chi(h)$ is proposed in the Section \ref{sec:rcn}.
   
\subsection{\label{sec:arbitraryDN} Susceptibility of an arbitrary directed network}

Many real-world networks can be represented by directed graphs. Well-known examples are the World Wide Web (WWW), neuronal and metabolic networks, and many other systems \cite{newman2010networks,dorogovtsev2022nature,newman2003structure}. In this subsection, using the field of observers,  we introduce the susceptibility of an arbitrary directed network, which have both directional and bidirectional edges. 

Directed networks have the following common structural properties. Any large directed network can be partitioned into qualitatively different subgraphs:
(i) the giant strongly connected component, and giant in- and out-components; (ii) finite directed components (tendrils and tubes); (iii) disconnected finite components \cite{broder2000graph,newman2001random,dorogovtsev2001giant,schwartz2002percolation,boguna2005generalized,timar2017mapping}. 
The giant components of the WWW  can be represented  by use of the ``bow-tie'' diagram \cite{broder2000graph}.  A rich hierarchical organization of tendril layers and tubes, that goes beyond the structure represented by the `bow-tie' diagram, was revealed in \cite{timar2017mapping}. 

In this section,  
we introduce the susceptibility of an arbitrary directed network and show that a strong increase of the susceptibility, when approaching the critical percolation point, is early-warning signal of the percolation phase transition and emergence or disappearance of a giant connected component. We show that it is a structural change of finite components, namely, tendrils, tubes and finite clusters,  that manifests critical phenomena in the percolation process.  

We chose at random a fraction $h$ of nodes in a directed networks. These nodes, which we call observers, form a set $\hat{C}_h$. For each node $i \in \hat{C}_h$, we find nodes, which can reach $i$ by following edge directness. We call this set of nodes as $\hat{T}_{in}(i)$. Note that $i \in \hat{T}_{in}(i)$. We also find a set $\hat{T}_{out}(i)$ of nodes, which we can reach starting from $i$ and then following edge directness. Alternatively, the set $\hat{T}_{out}(i)$ is the set of nodes, which can reach $i$ by following against the edge directness. We introduce a set 
\begin{equation}
 \hat{T}_{tot}(i) \equiv \hat{T}_{in}(i) \bigcup \hat{T}_{out}(i).
 \label{1sd}
\end{equation}
The following set, 
\begin{equation}
 \hat{T}_{tot} \equiv \bigcup_{i \in \hat{C}_h} \hat{T}_{tot}(i),
 \label{2sd}
\end{equation}
includes all nodes, which can reach at least one of the observers by following edge direction or against it.   The parameter 
\begin{equation}
S(h) \equiv \frac{1}{N} V(\hat{T}_{tot}) ,
 \label{3sd}
\end{equation}
is the fraction of the nodes belonging to the set $\hat{T}_{tot}$. The susceptibility of a directed network is 
\begin{equation}
\chi(h) = \frac{d S(h)}{d h}.
 \label{4sd}
\end{equation}
To prove that this definition agrees with the zero-field susceptibility introduced for directed networks in \cite{timar2017mapping}, we use the correlation function $C(i,j)$ for nodes $i$ and $j$ in the network $\hat{G}$. It has the following properties: (i) $C(i,i)=1$; (ii) $C(i,j)=1$ if there is directed pass along edge directness from $i$ to $j$. If a node $j$ belongs to the set $\hat{T}_{out}(i)$ then $C(i,j)=1$. If a node $j$ belongs to the set $\hat{T}_{in}(i)$ then $C(j,i)=1$. if $j \in \hat{T}_{in}(i)$, then $C(j,i)=1$. Note that this correlation function is the generalization of the correlation function of the Potts model on directed networks. Using this function we find the total number of nodes in the set $\hat{T}_{tot}$, Eq.~(\ref{2sd}). The relative fraction of the nodes in the considered network is     
\begin{equation}
S_h = \frac{1}{N} \sum_{i \in \hat{C}_h} \sum_{j \neq i} \Bigl[ C(j,i)+C(i,j)\Bigl] + \frac{1}{N}  \sum_{i\in \hat{C}_h, j=i} C(j,i), 
 \label{5sd}
\end{equation}
where the first summation over $j$ includes all nodes in the network apart $\hat{C}_h$. 

Let us take into account the organization of a directed network. The whole directed network $G$ is a uniting of sets: (i) the giant in-component, $\hat{I}$; (ii) the giant out-component, $\hat{O}$; (iii) the giant weakly connected component, $\hat{W}= \hat{I} \bigcup \hat{O}$; (iv) the strongly connected component is  $\hat{S}=\hat{I} \bigcap \hat{O}$  the tendrils and tubes, $\hat{T}$; (v) and the finite clusters, $\hat{F}$  \cite{broder2000graph,newman2001random,dorogovtsev2001giant,schwartz2002percolation,boguna2005generalized,timar2017mapping}:
\begin{equation}
 \hat{G} = \hat{W} \bigcup \hat{T} \bigcup \hat{F}.
 \label{6sd}
\end{equation}
In turn, $\hat{T}$ and  $\hat{F}$ are the unification of finite components: $\hat{T}=\bigcup_{\beta} \hat{T}_{\beta}$ and  $\hat{F}=\bigcup_{\alpha} \hat{F}_{\alpha}$, correspondingly. 
Taking into account that the observers are distributed uniformly at random over all of the components, we find 
\begin{eqnarray}
S(h) &=& \frac{1}{N}  \sum_{\alpha} f_{\alpha} [1-(1-h)^{f_{\alpha}}] 
\\
&+& \frac{1}{N}  \sum_{\beta} t_{\beta} [1-(1-h)^{f_{\beta}}] 
\\
&+& \frac{1}{N}  S_W [1-(1-h)^{S_{W}}], 
\label{7sd}
\end{eqnarray} 
 where $f_{\alpha}$ and $t_{\beta}$ are sizes of the finite components $F_{\alpha}$ and $T_{\beta}$. $S_W$ is the size of $\hat{W}$. In the thermodynamic limit $N  \gg 1$ at $h > 0$, the contribution of the giant week connected component $\hat{W}$ tends to zero. As a result, the susceptibility $\chi(h)$ is 
 \begin{equation}
\chi(h) = \frac{1}{N}  \sum_{\alpha} f_{\alpha}^2 (1-h)^{f_{\alpha}} + \frac{1}{N}  \sum_{\beta} t_{\beta}^2 (1-h)^{t_{\beta}}. 
\label{8sd}
\end{equation}
The zero-field susceptibility is 
\begin{equation}
\chi(0) = \frac{1}{N}  \sum_{\alpha} f_{\alpha}^2  + \frac{1}{N}  \sum_{\beta} t_{\beta}^2 , 
\label{9sd}
\end{equation}
that is $\chi(0)$ is completely determined by the finite components of the directed network. Numerical simulations of uncorrelated random directed networks in \cite{timar2021enhanced} showed that $\chi(0) \sim 1/|p-p_c|$.

It is important to note the use of  Eqs.~(\ref{1sd})-(\ref{4sd}) not only allows one to calculate the susceptibility of an arbitrary directed network, but also significantly reduces the computation time, since it does not require an analysis of the network structure. We just use information gathering by the observers distributed randomly over the network. In Section \ref{sec:rcn} we derive an analytical method to find $\chi (h)$ in uncorrelated tree-like directed networks.

\subsection{\label{sec:k-core} Susceptibility of the $k$-core}

In this subsection we introduce the susceptibility of the $k$-core. The $k$-core is the largest subgraph whose vertices have at least $k$ neighbors within this subgraph. An arbitrary network $\hat{G}$ can be represented as a set of successively enclosed giant $k$-cores \cite{dorogovtsev2006kcore}. Decreasing the occupation parameter $p$ destroys sequentially the $k$-cores starting from a maximum core and ending at the $2$-core.  At a given $p$, the $k$-core can be found by a sequential removal of nodes having degrees smaller than a given $k$, including the observers. At a critical parameter $p_c(k)$, the giant $k$-core collapses from a finite size to zero as a result of the mixed-order phase transition \cite{dorogovtsev2006kcore}.  Our aim is to introduce the susceptibility, which signals this phase transition. At this end, we choose at random a fraction $h$ of nodes, which form the field of observers in $\hat{G}$. At a given $p$ in the $k$-core, we find observers, which have degree $k$. They form a set $\hat{C}_h(k,p)$. Note, that nodes of the degree $k$  are called as corona nodes and they play important role in the $k$-core collapse.  For each node $i \in \hat{C}_h(k,p)$, we find all corona nodes, which are reachable from $i$ by following a path formed by other corona nodes. These corona nodes form a set $\hat{T_k(i)}$. The set of corona nodes reachable from the observers is                 
\begin{equation}
 \hat{T}_{kh}=\bigcup_{i \in \hat{C}_h(k,p)}\hat{T_k(i)}.
\label{1skc}
\end{equation}
The number of nodes in $ \hat{T}_{kh}$ is $v( \hat{T}_{kh})$. They occupy a fraction 
\begin{equation}
 S_k(p,h)=\frac{V(\hat{T}_{kh})}{NM(k)},
\label{2skc}
\end{equation}
where N$M(k)$ is the number of nodes in the $k$-core of the graph $\hat{G}$. The susceptibility of the $k$-core is 
\begin{equation}
\chi_k(p,h)= \frac{dS_k(p,h)}{dh}. 
\label{3skc}
\end{equation}
Let us show that at $h > 0$ the susceptibility has a maximum, when $p\rightarrow p_c(k)+0$ from above, and diverges at $h \rightarrow 0$. At this end, we take into account that corona clusters form a set of finite clusters $\hat{s}_{\alpha}$ in the $k$-core \cite{dorogovtsev2006kcore,goltsev2006kbootstrap}. We obtain
\begin{eqnarray}
 S_k(p,h) &=&  \frac{1}{NM(k)} \sum_{\alpha} s_{\alpha} [1-(1-h)^{s_{\alpha}}]
\\
&=& \frac{1}{NM(k)} \sum_{s} N_c(s) s [1-(1-h)^s],
\label{4skc}
\end{eqnarray}
where $N_c(s)$ is the number of corona clusters of size $s$. According to \cite{goltsev2006kbootstrap} in uncorrelated random networks, we have 
\begin{equation}
sN_c(s) = A s^{-3/2} \exp(-\frac{s}{s^{\ast}}).
\label{5skc}
\end{equation}
where $1/s^{\ast}(p) \sim (p-p_c(k)$. This gives the following asymptotic behavior:
\begin{equation}
\chi_k(p,h) \sim \frac{1}{\sqrt{h+1/s^{\ast}(p)}}.
\label{6skc}
\end{equation}
Thus,  we have  
\begin{eqnarray}
\chi_k(p,0) &\sim&  1/ \sqrt{p -p_c(k)},
\nonumber
\\
\chi_k(p_c(k),h) &\sim& 1/ \sqrt{h}.
\label{7skc}
\end{eqnarray}

\section{\label{sec:rcn} Susceptibility of uncorrelated random tree-like networks}

In this section we develop an analytical method to find the susceptibility of uncorrelated random tree-like networks. For this end, we use the method of generating functions \cite{newman2001random,dorogovtsev2008critical}.

\subsection{\label{sec:rcn} Susceptibility of uncorrelated random undirected tree-like networks}

Given a network $\hat{G}$, we chose at random a fraction $h$ of $N$ nodes and form a field of observers $\hat{C_h}$. Then, we remove nodes  at random with the probability $1-p$, including observers. The remaining observers monitor the remaining  part of the network. Let us find a fraction of remaining  nodes, from which one can reach at least one of the observers. We introduce a probability $Z$ that following a randomly chosen edge one can reach an observer. 
A self-consistent equation for $Z$ as function of $p$ and $h$ is determined by the following equation:
\begin{equation}
Z = ph + p(1-h)\sum_q P(q) \frac{q}{\langle q \rangle}[1 - (1-Z)^{q-1}] 
,
\label{1su}
\end{equation}
where $P(q)$ and $\langle q \rangle$ are the degree distribution and the mean degree. The first term is the probability that a randomly chosen edge leads directly to an observer. The second term in Eq.~(\ref{1su}) is the probability that a randomly chosen edge leads to a remaining node, which does not belong to  $\hat{C}_h$, but it has at least one edge, which leads to $\hat{C}_h$.
If we know $Z$, we can find the fraction of nodes, from which  one can reach $\hat{C}_h$,
\begin{equation}
S(h) = ph + p(1-h)\sum_q P(q) [1 - (1-Z)^{q}]
.
\label{2su}
\end{equation}
The first term is the fraction of remaining observers. The second term is the fraction of nodes, which do not belong to $\hat{C}_h$, but they have at least one edge, which leads to an observer. If the second moment $\langle q(q-1) \rangle$ of the degree distribution is finite, then according to the equations (\ref{1su}) and (\ref{2su}),  the percolation point is $p_c = \langle q \rangle / \langle q(q-1) \rangle$. 

Note that at $h > 0$, Eqs.~(\ref{1su})-(\ref{2su}) have a non-zero solution, $S(p,h)>0$, at any $ p > 0$, i.e., including a normal phase at $p < p_c$. This result is  due to the existence of finite clusters at $p>0$, see Eq.~(\ref{4sa}). This result is similar, for example, to an impact of a magnetic field, which produces a magnetic moment in ferromagnetic materials at temperature above the Curie temperature. At $h=0$, this set of equations (\ref{1su})-(\ref{2su}) has a singular solution corresponding to an appearance of the giant connected cluster with a size $S(p,0) > 0$ above a critical parameter $p_c$.     

An explicit analytical solution of the equations can be found for a regular random graph with a given coordination number $q$. This kind of networks is a simple and representative model of random uncorrelated networks with a finite second moment of degree distribution \cite{dorogovtsev2008critical}. In the case $q=3$, equations (\ref{1su}) and (\ref{2su}) take the following form:   
\begin{eqnarray}
Z &=& ph+p(1-h)\left[ 1 - (1 - Z)^2 \right]
,
\label{1r}
\\[3pt]
S(h) &=& ph+p(1-h)\left[ 1 - (1 - Z)^3 \right]
.
\label{2r}  
\end{eqnarray}
The analytical solution of Eq.~(\ref{1r}) is
\begin{equation}
Z = \frac{p-p_c -ph + \sqrt{p(1-p)h +(p-p_c)^2}}{p(1-h)}
.
\label{3su}
\end{equation}
Using Eqs.~(\ref{2r}) and (\ref{3su}), we find the susceptibility Eq.~(\ref{2sa}). Near the percolation point at $|p-p_c| \ll 1$, where $p_c =1/2$, at $h \ll 1$, we obtain
\begin{equation}
\chi(p,h) \approx \frac{3}{4\sqrt{h +4(p-p_c)^2}}
.
\label{4su}
\end{equation}
The susceptibility as a function of $p$ is a symmetric function near the maximum at $p=p_c$: $\max\chi(p,h)=\chi(p=p_c,h) \sim 1/\sqrt{h}$. At $h=0$, we have $\chi(p,h=0) \sim 1/|p-p_c|$ in agreement with Eqs.~(\ref{9sa}) and (\ref{10sa}). Equation (\ref{4su}) represents the universal mean-field behavior of the susceptibility near the continuous percolation phase transition with the critical exponent $\beta=1$. 

\subsection{\label{sec:sus-rdn} Susceptibility of uncorrelated random directed networks}

In this section, using the method of generating functions
\cite{newman2001random,dorogovtsev2001giant,schwartz2002percolation,boguna2005generalized,timar2017mapping}, we develop an analytical method to find the susceptibility of uncorrelated random directed networks, which have a local tree-like structure.
We choose at random a fraction $h$ of nodes in an uncorrelated random directed network. The chosen nodes are observers. Then, nodes, including the observers, are removed with the probability $1-p$. Following Section~\ref{sec:arbitraryDN}, we aim to find the fraction of remaining nodes which can reach remaining observers by following along or against the edge directness.  For this end, we introduce the probabilities $Z_{in}(h)$ and $Z_{out}(h)$.  $Z_{in}(h)$ is the probability that following along the edge directness of a randomly chosen edge we can reach an observer. $Z_{out}(h)$ is the probability that following against the edge directness of a randomly chosen edge we can reach an observer.  The probabilities are a solution of the following two equations:
\begin{eqnarray}
Z_{in}(h) &{=}& hp +(1{-}h)p \sum_{k_{in}, k_o} \frac{k_{in}}{\langle  k_{in} \rangle} P(k_{in}, k_{o})
\nonumber
\\[3pt]
&{\times}&  \Bigl[ 1{-} (1{-}Z_{in}(h))^{k_o} \Bigl],
\label{1rdn}
\\[3pt]
Z_{out}(h) &=& hp + (1{-}h)p \sum_{k_{in}, k_o} \frac{k_{o}}{\langle  k_{o} \rangle} P(k_{in}, k_{o})
\nonumber
\\[3pt]
&{\times}&  \Bigl[ 1{-} (1{-}Z_{out}(h))^{k_{in}} \Bigl].
\label{2rdn}
\end{eqnarray}
Here, $P(k_{in}, k_{o})$ is the degree distribution function of incoming and outgoing edges. $\langle  k_{in} \rangle$ and $\langle  k_{o} \rangle$ are the mean degrees of incoming  and outgoing edges, $\langle  k_{in} \rangle=\langle  k_{o} \rangle$. The first terms in Eqs.~(\ref{1rdn}) and (\ref{2rdn}) are the probabilities that a randomly chosen edge is an  incoming or outgoing edge of an observer, respectively. The seconds terms are probabilities that a randomly chosen edge  leads to an observer if we follows along or against its edge directness, respectively.  
The fraction $S(h)$ of nodes, which can reach the observers, moving along or against edge directness equals 
\begin{eqnarray}
S(h)&{=}&ph+(1{-}h)p \sum_{k_{in}, k_o} P(k_{in}, k_{o}) \Bigr([1{-}(1{-}Z_{in}(h))^{k_{in}}]
\nonumber
\\
&+& [1-(1{-}Z_{out}(h))^{k_o}] \Bigl).
\label{3rdn}
\end{eqnarray}
According to the equations Eqs.~(\ref{1rdn}) and (\ref{2rdn}), at $h=0$ the strongly connected component emerges at the point
\begin{equation}
p_c = \frac{\langle  k_{o} \rangle}{\langle k_{in} k_{o} \rangle}. 
\label{4rdn}
\end{equation}
In the case of a symmetrical distribution $P(k_{in}, k_o) = P(k_o, k_{in})$, solving Eqs.~(\ref{1rdn}) and (\ref{2rdn}) at $h \ll 1$ and  $|p-p_c| \ll p_c$, we find
\begin{equation}
\chi(p,h) = \frac{2p^2 \langle  k_{o} \rangle}{\sqrt{Ah + (1-p/p_c)^2}}. 
\label{5rdn}
\end{equation}
where $A = 2p^2 \langle k_{in} k_{o} (k_o -1) \rangle/ \langle k_{in} \rangle$. The susceptibility
$\chi(p,h)$ is a symmetric function of $p$ near $p_c$. It reaches a maximum, $\chi(p_c,h) \sim 1/\sqrt{h}$, at the critical point. At $h \rightarrow 0$, we find $\chi(p,h=0) \sim 1/|p-p_c|$. Equation (\ref{5rdn}) is equivalent to Eq.~(\ref{4su}) for random undirected networks.

\subsection{\label{sec: trees} Susceptibility of uncorrelated random networks with dependency links.}

In this section we find the susceptibility of uncorrelated random undirected complex networks with dependency links introduced in \cite{parshani2011critical}. In this kind of networks, apart usual connectivity edges, which link nodes, there are also so called dependency links, which provide an additional condition on the mutual existence of pairs of nodes connected by dependency links. If one of the node in a pair is removed, then the other node in the pair also is removed. If the nodes in a pair belong to different clusters then they must be removed. Only if nodes in a pair belong to the same cluster, then they present in the network. It was demonstrated in \cite{parshani2011critical,bashan2011percolation,bashan2011combined,lin2017robustness,timar2021enhanced} that the presence of dependency links results in a rich phase diagram of the percolation process with a continuous phase transition, the tricritical point, and a discontinuous mixed-order phase transition. We aim to find the critical behavior of the susceptibility near the phase transitions. 
      
\subsubsection{\label{sec: toymodel} Exactly solvable model with dependency links}

We consider a $q$-regular random network with the coordination
number $q = 3$ and dependency links. Each
node has with a probability $r$ a single dependency link
to a randomly selected node. First we show that depending
on $r$, this model demonstrates an ordinary continuous
percolation phase transition, a tricritical point, and
a mixed-order phase transition.

\begin{figure}[b]
\begin{center}
\includegraphics[width=0.47\textwidth]{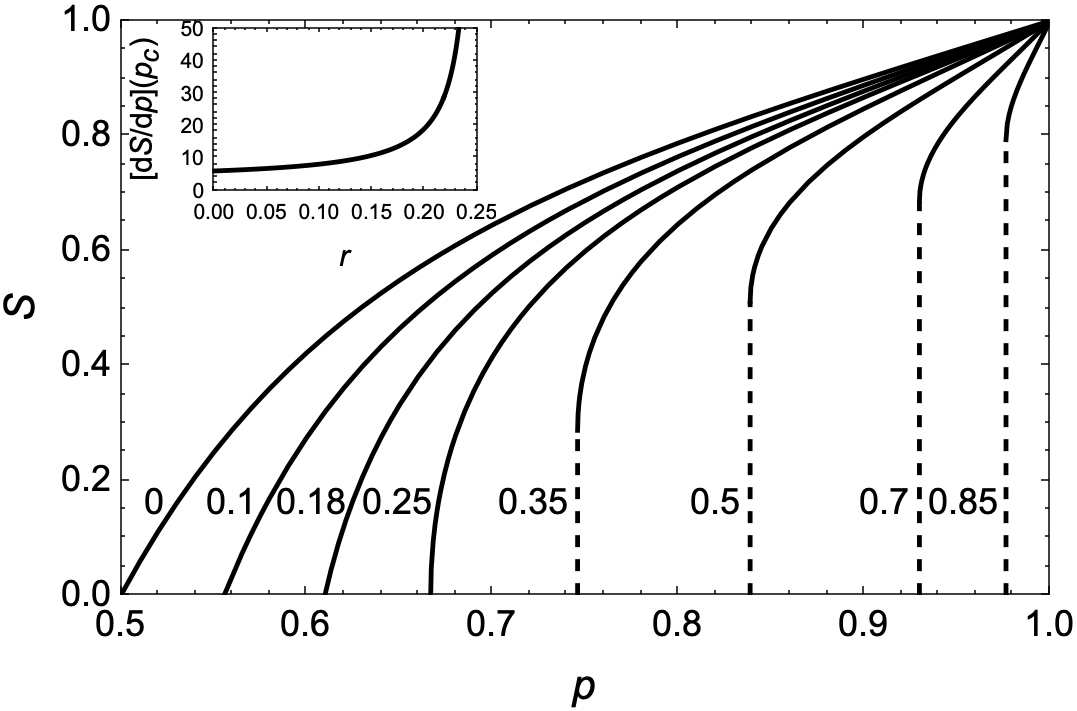}
\caption{The size of the giant cluster $S$ vs. $p$ in a $3$-regular random graph with the probability $r$ for nodes to have a dependency link. The curves from left to right: $r = 0$, $0.1$, $0.18$, $0.25$, $0.35$, $0.5$, $0.7$, $0.85$. For $r=1$, the giant connected percolation exists only at $p=1$, where $S=1$. The phase transition is of second order for $r < 1/4$, when $S \propto (p-p_c(r))$ near the critical parameter $p_c(r)$. The curve $r=0.25$ corresponds to the tricritical point. The phase transition is hybrid for $r > 1/4$. The jump of $S$ is shown by the dashed line. At $p=p_c(r)$, the singularity of $S$ is square-root, including the case of $r=1/4$. The insert shows the derivative $(dS/dp)|_{p=p_c(r)}$ at the critical point of the continuous percolation transition. 
},
\label{f1}
\end{center}
\end{figure}


First we consider the case $h=0$. We remove nodes with the probability $1-p$. Accounting that two nodes, which are interconnected by a dependency link, remain if they belong to the giant connected component, we renormalize the parameter $p$ in Eq.~(\ref{1r}) and (\ref{2r}) as follow:
\begin{equation}
p \to p - pr(1 - S)
.
\label{30}
\end{equation}
Here, the second term is the probability that a node is removed if its partner, to which it is connected by the dependency link, does not belong to the giant connected component.  We obtain the following equations for $Z$ and $S$:
\begin{eqnarray}
Z &=& \frac{p(1 - r)Z(2 - Z)}{1 - p r Z(3 - 3Z + Z^2)}
,
\label{42}
\\[3pt]
S &=& \frac{p(1 - r)[1 - (1 - Z)^3]}{1 - pr[1 - (1 - Z)^3]}
.
\label{44}   
\end{eqnarray}
These equations give a cubic equation for $S$:
\begin{eqnarray}
S = p(1 + rS - r) \left[ 2 - \frac{1}{p^3(1 + rS - r)^3} \right.
\nonumber
\\
 +\left. \!\frac{3}{p^2(1 + rS - r)^2} - \frac{3}{p(1 + rS - r)} \right]
.
\label{40}   
\end{eqnarray}
It has three roots, which can be written explicitly. For small $Z$, Eq.~(\ref{42}) reads
\begin{eqnarray}
[1 - 2p(1 - r)]Z &+& p(1 - r)(1 - 6pr)Z^2
\nonumber
\\[3pt]
&+& 9p^2 r(1 - r)(1 - 2 pr) Z^3 = 0
.
\label{46}    
\end{eqnarray}
Equating the coefficient of the $Z$ term to zero provides the continuous transition line: $1 - 2p_c(1 - r) = 0$. Equating both coefficients of the $Z$ and $Z^2$ terms to zero provides the special point $(r^\ast,p^\ast)$ corresponding to the crossing
of two lines: $1 - 2p(1 - r) = 0$ and $1 - 6pr = 0$ on the $(r,p)$ plane. Thus, $(r^\ast,p^\ast)=(1/4,2/3)$. Figure~\ref{f1} shows the resulting curves $S(p)$ for various $r$. For $r=1$, the giant connected component ($S=1$) exists only at $p=1$. For $r<1/4$, the phase transition is second-order at the critical point
\begin{equation}
p_{c}(r)= \frac{1}{2(1 - r)}
.
\label{50}    
\end{equation}
Near the critical point at $p - p_{c}(r) \ll 1$, we get
\begin{eqnarray}
Z &\cong& \frac{4(1 - r)^2}{1 - 4r}\,(p - p_{c}(r))
,
\\
\label{370}
S&\cong& 6\,\frac{(1 - r)^2}{1 - 4r} (p-p_c(r))
.
\label{93}    
\end{eqnarray}
Note that the derivative $(dS/dp)|_{p=p_c}$, i.e., the slope of $S$, increases, when $r$ tends to $r^\ast=1/4$, see the insert in Fig.~\ref{f1}. It diverges at $r= r^\ast$ that signals a transition to a tricritical point. Notably, at the special point $r^{\ast}$, the singularity
turns out to be square-root that corresponds to the critical exponent $\beta=1/2$:
\begin{equation}
S \cong \frac{3\sqrt{3}}{2} \sqrt{p - p_c(r^\ast)}
,
\label{60}    
\end{equation}
where $p_c(r^\ast)= 2/3$. 

\begin{figure}[t]
\begin{center}
\includegraphics[width=0.47\textwidth]{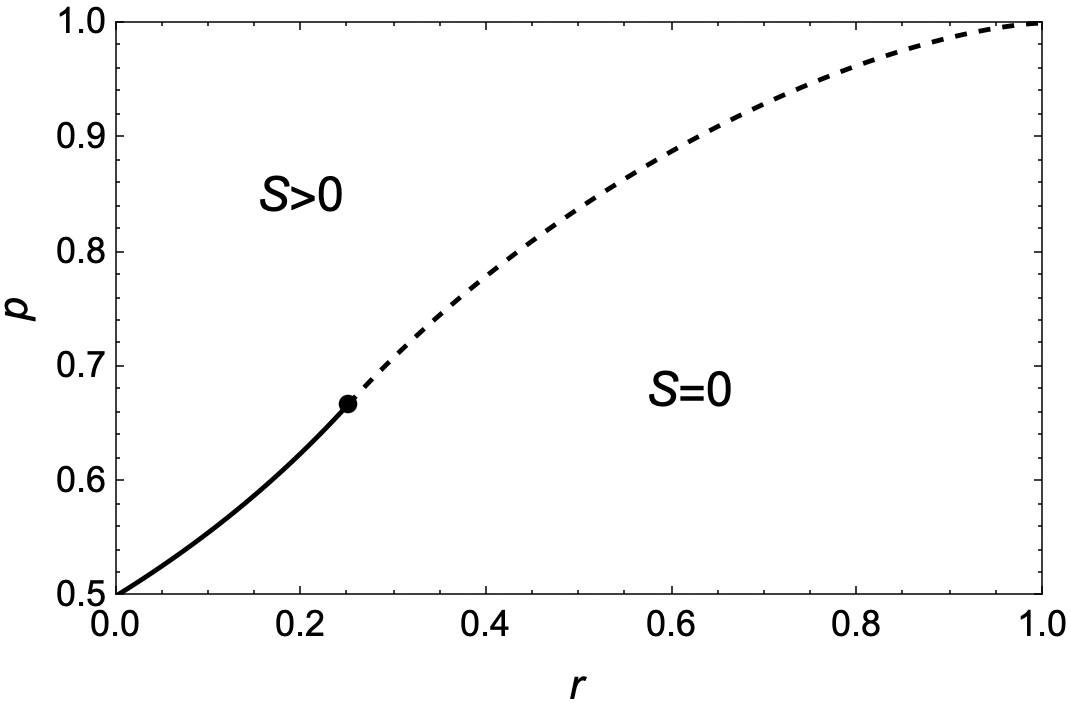}
\caption{Phase diagram on the $(r,p)$-plane for a $3$-regular random graph with dependency links. The solid curve represents the critical points of the continuous percolation transitions with the critical exponent $\beta =1$, see Eq.~(\ref{50}). The black dot shows the tricritical point  with $\beta =1/2$. The dashed curve shows the critical curve of the mixed order phase transition. The region above the critical curve is the region with the giant connected component, i.e., $S>0$. The region below the curve is the region with $S=0$.  
}
\label{f2}
\end{center}
\end{figure}

For $r>1/4$, the phase transition is mixed-order with a square-root singularity at the critical point
\begin{equation}
p_h(r)=p_c(r) - \frac{64}{27}(r- r^{\ast})^2
.
\label{105}
\end{equation}
Solving the Eqs.~(\ref{42}) and (\ref{44}) numerically, we obtained the $S(p)$ versus $p$, see Fig.~\ref{f1}, and the phase diagram on the $(r,p)$-plane, see Fig.~\ref{f2}. A similar phase diagram was observed within different models in \cite{parshani2011critical,timar2021enhanced}.

\begin{figure}[t]
\begin{center}
\includegraphics[width=0.47\textwidth]{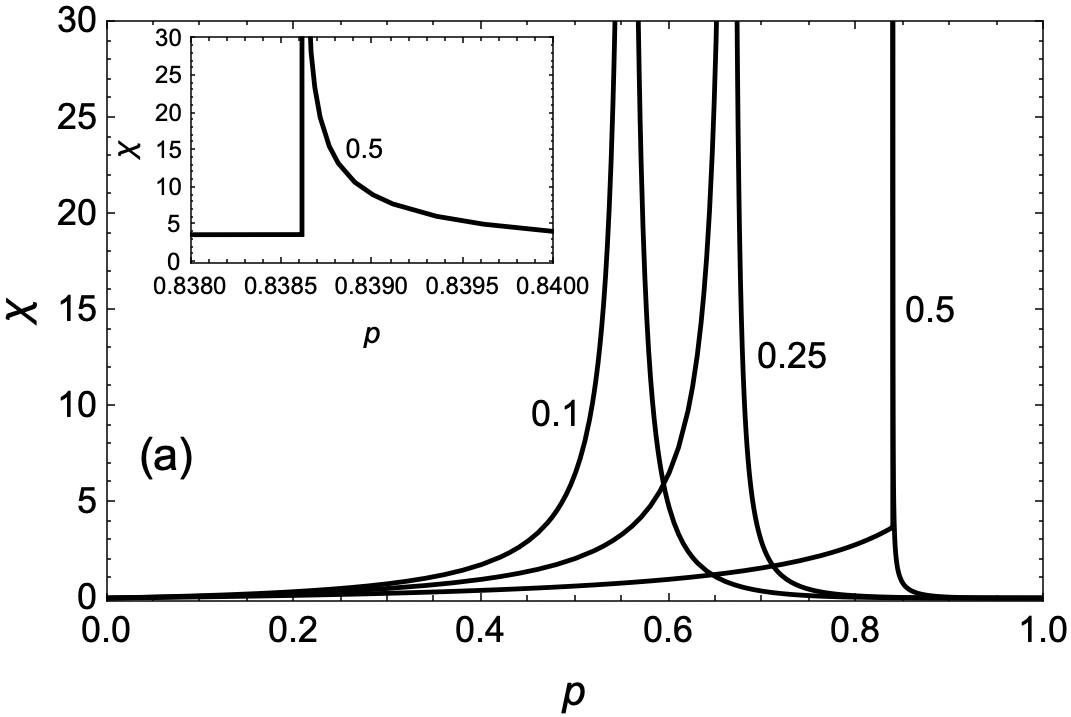}
\\
\includegraphics[width=0.47\textwidth]{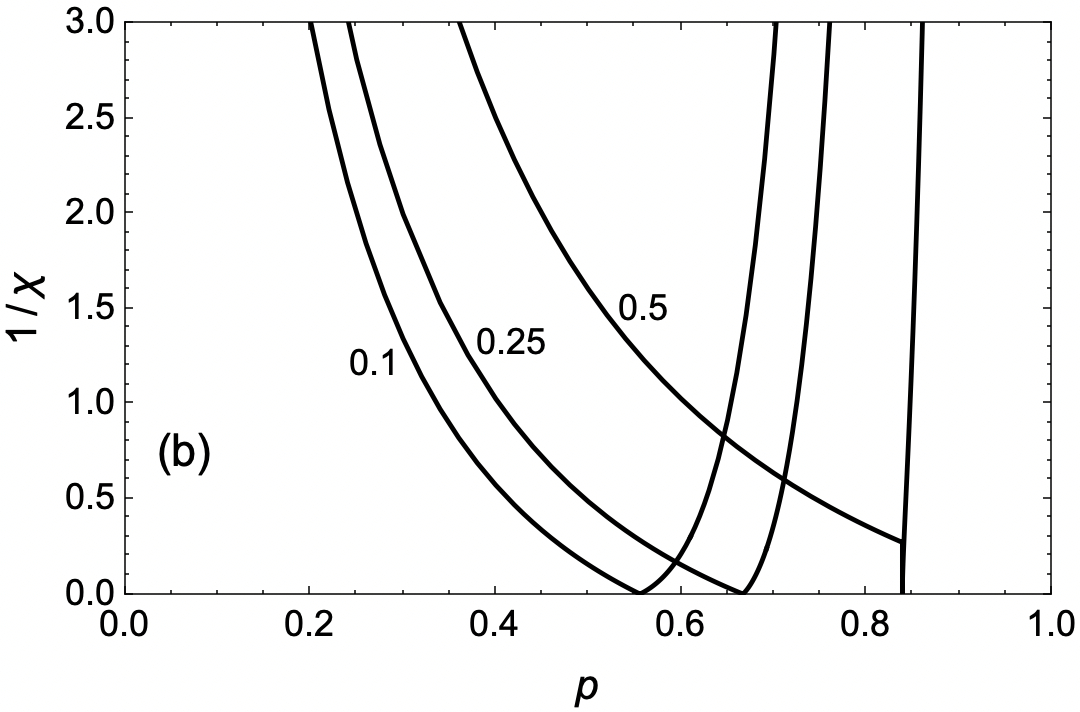}
\caption{(a) The susceptibility  $\chi$ vs. $p$ in a $3$-regular random graph with dependency links for different values of the probability $r$ to have a dependency link: (i) $r = 0.1$, the ordinary percolation transition with the critical exponent $\beta=1$; (ii) $r=r^\ast=0.25$, the tricritical point with $\beta=1/2$; (iii) $r=0.5$, the mixed order phase transition. The insert represents the zoom of $\chi$ at $r=0.5$. (b) The corresponding $1/\chi$ vs. $p$.
}
\label{f3}
\end{center}
\end{figure}


Let us find the susceptibility $\chi(p,h)$ of the $3$-regular random graph with dependency links. Following the Sections \ref{sec:arbitrary} and \ref{sec:rcn}, we choose at random a fraction $h$ of nodes and form the field of observers $\hat{C}_h$. Then we remove at random the fraction $1-p$ of nodes including the observers. Some of the removed nodes may have dependency links, so their partners also must be removed, and so on. We aim to find the fraction $S(h)$ of the remaining nodes which can reach observers by following connectivity edges.  Substituting Eq.~(\ref{30}) into Eqs.~(\ref{1r}) and (\ref{2r}), we get
\begin{eqnarray}
Z &=& p(1 + rS - r) \left\{h + (1-h)\left[ 1 - (1 - Z)^2 \right] \right\}
,
\label{320}
\\[3pt]
S &=& p(1 + rS - r) \left\{ h + (1-h)\left[ 1 - (1 - Z)^3 \right] \right\}
,
\label{330}  
\end{eqnarray}
and hence
\begin{eqnarray}
S &=&
\frac{ p(1 - r) \left\{ h + (1-h)\left[ 1 - (1 - Z)^3\right] \right\} }{ 1 - pr \left\{ h + (1 - h) \left[ 1 - (1 - Z)^3 \right] \right\} }
,
\label{340}
\\[3pt]
Z &=& p
\nonumber
\\[3pt]
&\times&
\left\{ 1 + r\,\frac{ p(1 - r) \left\{ h + (1-h)\left[ 1 - (1 - Z)^3\right] \right\} }{ 1 - pr \left\{ h + (1 - h) \left[ 1 - (1 - Z)^3 \right] \right\} } - r \right\}
\nonumber
\\[3pt]
&\times&
\left\{ h + (1-h)\left[ 1 - (1 - Z)^2 \right] \right\}
.
\label{350}    
\end{eqnarray}
Applying $(d/dh)|_{h=0}$ to Eqs.~(\ref{340}) and (\ref{350}), we get a (cumbersome) expression of susceptibility in terms of $Z$. Substituting $Z=0$ into this expression, we obtain the zero-field susceptibility $\chi$ in the normal phase, $p<p_c(r)$:
\begin{equation}
\chi(p,0) = p(1 - r) + \frac{3}{2}\ \frac{p^2 (1 - r)}{p_c(r) - p}
\cong \frac{3}{8(1-r)}\ \frac{1}{p_c(r) - p}
.
\label{360}    
\end{equation}
Using Eq.~(\ref{370})), in the ordered phase, $p>p_c(r)$, we obtain
\begin{equation}
\chi(p,0) \cong \frac{3}{8(1-r)}\, \frac{1}{p - p_{c}(r)}
,
\label{380}   
\end{equation}
which is symmetrical to the susceptibility in the normal phase, Eq.~(\ref{360}).

At the tricritical point $r=r^{\ast}=1/4$, the zero field susceptibility in normal phase is
\begin{equation}
\chi(p,0) \cong \frac{1}{2}\ \frac{1}{p_{c}(r^\ast) - p}
.
\label{390}    
\end{equation}
Using Eq.~(\ref{60}), in phase with the giant connected component, $p>p_c (r^\ast)$ we get
\begin{equation}
\chi(p,0) \cong \frac{1}{4}\, \frac{1}{p - p_c (r^\ast)}
,
\label{410}    
\end{equation}
One can see that the critical behavior of the zero-field susceptibility is asymmetrical above and below  $p_c (r^\ast)$ in contrast to Eqs.~(\ref{360}) and (\ref{380}).
Explicit results for $\chi(p,0)$ vs. $p$ at various $r$ are shown in Figure \ref{f3}(a). One can see
the continuous percolation transitions with the critical
exponents $\beta =1 $ and $\beta =2 $, see the curve corresponding
to $r = 0.1$ and $r = 0.25$, respectively. Figure \ref{f3}(b) shows
$1/\chi(p,0)$ vs. $p$ for various $r$.

Finally, for $r> 1/4$, the phase transition is the mixed-order transition. Above the critical point $p_h(r)$, Eq.~(\ref{105}), the giant connected component appears in a jump. At $p-p_h(r) \ll 1$ we find
\begin{equation}
\chi(p,0) \sim \frac{1}{\sqrt{p - p_h(r)}}
.
\label{420}    
\end{equation}
At $r > r^{\ast}$, when $p$ below the critical point $p_h(r)$ of the mixed-order transition, i.e., in the normal phase, the zero-field
susceptibility is finite,
\begin{equation}
\chi(p_h(r)-0,0) \cong \frac{27}{128}\, (r - r^{\ast})^{-2}
.
\label{420}    
\end{equation}
Note that $\chi(p_h(r)-0,0)$ diverges when $r \rightarrow r^{\ast}$. 
This signals approaching to the tricritical point.
Thus, the zero-field susceptibility $\chi(p, h = 0)$ demonstrates an asymmetric behavior as a function of $p$ near
$p_h(r)$, see the insert in Fig.~\ref{f3}(a).


\subsubsection{\label{arbitrary} Susceptibility for arbitrary distribution functions of connectivity and dependency links}

We consider an uncorrelated random network with the degree distribution $P_c(q)$ of connectivity edges and the distribution $P_d(m)$ of dependency links. We choose at random a fraction $h$ of nodes and form a set of observers $\hat{C}_h$. With the probability $1-p$, nodes are removed. Our aim is to find the fraction $S(p,h)$ of remaining active nodes, which have at least one connectivity edge leading to active observers.  The following equations describe percolation process:
\begin{eqnarray}
Z &=&hp \sum_m P_d(m) Y^{\hspace{0.7pt} m}
\nonumber
\\[3pt]
&{+}& (1{-}h)p \sum_m P_d(m) Y^{\hspace{0.7pt}m} \, \sum_q \frac{qP(q)}{\langle q\rangle} \left[ 1 {-} (1 {-} Z)^{q{-}1} \right]
,
\nonumber
\\[3pt]
\label{500}
\\[3pt]
Y &=& hp \sum_m \frac{P_d(m)}{\langle m \rangle}\, Y^{\hspace{0.7pt}m-1}
\nonumber
\\[3pt]
&{+}& (1{-}h)p \sum_m \frac{P_d(m)}{\langle m \rangle}\, Y^{\hspace{0.7pt}m{-}1} \, \sum_q P(q)\! \left[ 1 {-} (1 {-} Z)^q \right]
,
\nonumber
\\[3pt]
\label{510}
\\[3pt]
S(h)&=& hp \sum_m  P_d(m) Y^{\hspace{0.7pt}m}
\nonumber
\\[3pt]
&{+}&(1{-}h)p \sum_m P_d(m) Y^{m} \, \sum_q P(q)\! \left[ 1 {-} (1 {-} Z)^q \right]
,
\nonumber
\\[3pt]
\label{520}  
\end{eqnarray}
Here, $Z$ is the probability that a randomly chosen connectivity edge leads to an active observer. $Y$ is the probability that a randomly chosen dependency links leads to an active observer.
The first terms in the Eqs.~(\ref{500}) and (\ref{510}) are the probabilities that a randomly chosen connectivity edge in Eq. (\ref{500}) and a dependency link in Eq. (\ref{510}) lead directly to an active observer. The second terms are the probabilities that a randomly chosen connectivity and dependency edges, respectively, lead to an active node, which is not the observer but has at least one connectivity edge leading to an active observer.
The equation (\ref{520}) is the analytical representation of  Eq.~(\ref{0sa}).  Solving  Eqs.~(\ref{500})-(\ref{520}) at $h=0$ , we find the critical point $p_c$ of the continuous phase transition and the size $S(0)$ of the giant connected component near $p_c$: 
\begin{eqnarray}
p_c & =& \frac{1}{P_d(0)}\,\frac{\langle q\rangle}{\langle  q(q-1) \rangle}
,
\label{550}
\\[3pt]
S(0) &{\simeq}& \frac{p{-}p_{c}}{p} \frac{2\langle m \rangle {P_d(0)}^2 \langle q(q{-}1) \rangle} {2{P_d(1)}^2{\langle q \rangle ^2{-}\langle m \rangle}{P_d(0)}^2 \langle q(q{-}1)(q{-}2) \rangle }.
\nonumber
\\[3pt]
\label{8g}   
\end{eqnarray}
If $h$ is nonzero but small, e.i., $h \ll 1$, then at the critical point $p=p_c$ we find the universal behavior: $S(h) \sim \sqrt{h}$ and $\chi \sim 1/\sqrt{h}$.   
If the distributions functions of the connectivity and dependency edges satisfy the equality,
\begin{equation}
2{P_d(1)}^2{\langle q \rangle ^2{-}\langle m \rangle}{P_d(0)}^2 \langle q(q{-}1)(q{-}2) \rangle=0,     
\end{equation}
then the slop $dS(0)/dp \mid_{p=p_c}$ diverges that means that the model is at the tricritical point, which corresponds to the continuous transition with $\beta =1/2$ and the susceptibility Eqs.~(\ref{390}) and (\ref{410}). When $P_d(1) = 0$, the mixed-order phase transition is impossible. 


\section{\label{sec:landau} Susceptibility within the Landau theory.}

In this section we consider the phenomenological Landau mean-field theory, which reproduces the critical behavior found above for undirected and directed networks and the phase diagram  in Figs.~\ref{f1}-\ref{f3} for the complex network with dependency edges. The aim is to demonstrate the universality of critical behavior of the susceptibility in complex networks. 

Let us study a system whose free energy $\Phi(x,p,r)$ is an analytical function of the order parameter $x$ and two control parameters $p$ and $r$. The function $\Phi(x,p,r)$ is represented as a series in powers of $x$,
\begin{equation}
\Phi(x,p,r)=\sum_{n=1}^{\infty}\frac{1}{n!}f_{n}(p,r)x^n.
\label{1p}
\end{equation}
In a general case,
the coefficients $f_{n}(p,r)$ are functions of the control parameters $p$ and $r$.
Assuming $x\ll 1$, in this expansion, we consider only first four terms:
\begin{eqnarray}
f_1&=&-h
,
\nonumber
\\[3pt]
f_2(p,r)&=& a_{2}(p_c -p)
,
\nonumber
\\[3pt]
f_{3}(p,r)&=&a_3 (r^{\ast}-r)
,
\nonumber
\\[3pt]
f_{4}(p,r)&=& f_4 
,
\label{2p}
\end{eqnarray}
where $h$ is a field conjugated to the order parameter $x$. We assume that  the coefficients $a_2$, $a_3$, and $f_4$ are positive.  Note that the coefficient $f_{3}(p,r)$ is a function of $r$. It changes its sign at $r=r^{\ast}$. $f_{3}(p,r)$ is positive at $r < r^{\ast}$, negative at $r > r^{\ast}$, and zero, $f_{3}(r^{\ast})=0$, at $r=r^{\ast}$. 

\subsection{Order parameter and phase diagram}

Let us find the phase diagram of the system on $(p,r)$ plane. At first, we consider the case  $r < r^{*}$.  The order parameter $x$ as a function of the control parameters $p$, $r$, and $h$  is determined by an equation,
\begin{equation}
\frac{\partial \Phi(x,p,r)}{\partial x} {=}{-}h{+}a_{2}(p_c{-}p) x{+}\frac{1}{2}a_3 (r^{*}{-}r) x^2 {+}\frac{1}{6}f_{4}x^3{=}0
.
\label{3p} 
\end{equation}
Solving this equation in the leading order $O(x^2)$ gives 
\begin{equation}
x(h) \simeq \frac{\sqrt{2h a_3 (r^{*}-r)+a_{2}^2(p-p_c)^2} \pm a_{2}(p-p_c)}{a_3 (r^{*}-r)},                     
\label{4ph}   
\end{equation}
where the sign $\pm$ corresponds  to $p>p_c$ and $p<p_c$, respectively.  One can see that when increasing $p$ at $h=0$  and $r< r^{\ast}$,  the system undergoes a continuous phase transition at $p=p_c$. Namely,  $x=0$ at $p\leq p_c$ and
\begin{equation}
x \simeq \frac{2a_{2}}{a_{3}(r^{\ast}-r)} (p-p_c)
\label{4p}   
\end{equation}
at $p > p_c$.  This transition has the critical exponent $\beta=1$ and corresponds to the percolation transition. Note that the slope $dx/dp$ diverges when $r \rightarrow r^{*}$. This signals the change of the critical behavior. 

At the $r=r^{*}$ and $h=0$, equation (\ref{3p}) gives the second-order phase transition with $\beta=1/2$. The order parameter is
\begin{equation}
x \simeq \sqrt{\frac{3a_2}{f_4}}\sqrt{p-p_c}
,
\label{5p}   
\end{equation}
at $p > p_c$. This is the tricritical point. 

At the $r>r^{\ast}$ the system demonstrates the first-order phase transition. The order parameter is
\begin{equation}
x \simeq \frac{3a_3 (r-r^{\ast})}{2f_4}+\sqrt{\frac{6a_2}{f_4}}\sqrt{p-p_h(r)}
,
\label{6p}  
\end{equation}
where the critical point $p_h(r)$ is
\begin{equation}
p_h(r) = p_c - \frac{3a_{3}^2(r-r^{\ast})^2}{8a_{2} f_4}
.
\label{7p}  
\end{equation}
Interestingly, that this equation is similar to the Eq.~(\ref{105}). The difference between Eq.~(\ref{105}) and Eq.~(\ref{7p}) is that here $p_c$ is a constant.
The jump of $x$ at $p=p_h(r)+0$ equals $x(p=p_h(r)+0)=3a_{3}(r-r^{\ast})/(2 f_4)$. It increases linearly when increasing $r$ above $r^{\ast}$.

\subsection{Susceptibility}

Using the phenomenological Landau theory, we find the susceptibility $\chi=dx/dh$.  Differentiating Eq.~(\ref{3p}) over $h$ gives us the susceptibility, $\chi(p,r,h)$, as a function of $p$, $r$ and $h$,
\begin{equation}
\chi(p,h,r) = \frac{1}{f_2 +f_3 x + f_4 x^2/2}
.
\label{8p}   
\end{equation}
At $r<r^{\ast}$ using Eq.~(\ref{4ph}), we find the susceptibility $\chi(p,h,r)$ near the continuous transition with $\beta =1$,
\begin{equation}
\chi(p,h,r) = \frac{1}{\sqrt{2h a_3 (r^{\ast}-r)+ a_2 (p-p_c)^2}}
.
\label{9p}    
\end{equation}
$\chi(p,h,r)$ has a maximum at the critical point $p = p_c$: $\max \chi(p,h,r) =\chi(p=p_c,h,r) \sim 1/\sqrt{h}$. The $\chi(p, h, r)$ is
a symmetric function of $p$ near $p_c$. This behavior of $\chi(p,h,r)$ has been discussed above for complex networks, see Eqs.~(\ref{9sa}) and (\ref{4su}).

If $r=r^{\ast}$, then the system is at the tricritical point  with the critical index $\beta = 1/2$, see  Eq.~(\ref{5p}). The zero-field susceptibility is
\begin{eqnarray}
\chi(p,0,r^{\ast}) = \frac{1}{a_2 (p_c-p)}
,
p<p_c
,
\nonumber
\\[3pt]
\chi(p,0,r^{\ast}) = \frac{1}{2a_2 (p-p_c)}
,
p>p_c
.
\label{10p}
\end{eqnarray}
The susceptibility demonstrates asymmetric behavior versus $p$ near $p_c$ in agreement with Eqs.~(\ref{390}) and (\ref{410}).
Solving Eq.~(\ref{3p}) at $p=p_c$ gives $x(p_c,h,r^{\ast})=(6h/f_4)^{1/3}$. The susceptibility $\chi(p,h,r)$ as a function of $p$ reaches a maximum at $p=p_c$, 
\begin{equation}
\max \chi(p,h,r) =\chi(p=p_c,h,r^{\ast}) \sim h^{2/3}
.
\label{9ph}    
\end{equation}

To find the susceptibility at the tricritical point in a scale free complex network with the degree exponent $3 < \gamma \leq 5$, we use the following free energy \cite{goltsev2003critical}:  
\begin{equation}
\Phi(x,p)=-hx +\frac{1}{2}a_2 (p_c -p) x^2 + Bx^{\gamma -1}.
\label{13p}
\end{equation}
At the critical point $p=p_c$, the susceptibility is 
\begin{equation}
\chi(p_c,h) \sim \frac{1}{h^{(\gamma - 3)/(\gamma - 2)}}
,
\label{12p} 
\end{equation}
Thus, the dependence of $\chi(p_c,h)$ on $h$ at the critical point is determined by the network topology. At $\gamma >5$, the upper meaningful term in the free energy Eq. (\ref{13p}) is $O(x^4)$ and the critical behavior is given by Eq.~(\ref{9ph}).

Using Eq.~(\ref{6p}), we find $\chi(p,r,0)$ near the hybrid transition at $p=p_h(r)$,
\begin{eqnarray}
\chi(p,0,r) &=& \frac{1}{a_2 (p_c(r)-p)}
, p<p_h(r)
,
\nonumber
\\[3pt]
\chi(p,0,r)&{=}&\Bigl[2a_{2}(p{-}p_{h}(r))
\nonumber
\\[3pt]
&{+}&\frac{1}{2}a_{3}(r{-}r^{\ast})[6a_{2}(p{-}p_h(r))/f_4]^{1/2}\Bigr]^{-1}
,
\nonumber
\\[3pt]
p>p_h(r)
.
\nonumber
\\[3pt]
\label{11p}
\end{eqnarray}
One can see that $\chi(p,0)$ is finite,
\begin{equation}
\chi(p,0,r)|_{p=p_h(r) -0}= \frac{8f_4}{3a_3 (r-r^{\ast})^2}
,
\label{12p} 
\end{equation}
When $p \rightarrow p_h(r)+0$ in the ordered phase, the susceptibility
$\chi(p, 0, r)$ diverges as
\begin{equation}
\chi(p,0,r) \sim  \frac{1}{\sqrt{p-p_h(r)}} 
.
\label{13p} 
\end{equation}
At $h \neq 0$ and $p=p_h(r)$, it reaches a maximum value,
\begin{equation}
\chi(p_h(r),h,r) = \frac{2}{\sqrt{h (r^{\ast} - r) a_3}}
,
\label{14p} 
\end{equation}
The $k$-core susceptibility demonstrate the same critical
behavior, see Eq.~(\ref{7skc}).
\begin{table}
\caption{\label{tab:table1}Critical behavior of the susceptibility $\chi (p,h)$ versus the occupation probability $p$ and the field strength $h$ near
the critical point $p_c$ in complex networks for continuous percolation phase transitions with the critical exponent $\beta$ and
mixed-order transitions. Here, $\max{\chi (p,h)} = \chi (p_c,h)$.}
\begin{ruledtabular}
\begin{tabular}{cccc}
 & disordered state& ordered state& \\
  & & & \\
 &$\chi (p,0)$ &$\chi (p,0)$ & $\chi (p_c,h)$\\
\hline
 & & & \\
$\beta=1$ &$(p_c - p)^{-1}$ &$(p - p_c)^{-1}$ & $h^{-1/2}$\\
 & & & \\
 $\beta=1/2$ &$(p_c - p)^{-1}$ &$\frac{1}{2}(p - p_c)^{-1}$ & $h^{-2/3}$\\
  & & & \\
mixed-order & constant & $(p - p_c)^{-1/2}$ & $h^{-1/2}$ \\
  & & & \\
$k$-core  & 0 &  $(p - p_c)^{-1/2}$ &  $h^{-1/2}$  
\end{tabular}
\end{ruledtabular}
\end{table}

The critical behavior of the order parameter and susceptibility,
which we obtained by use of the phenomenological
Landau theory with the free energy Eq.~(\ref{1p}) and
(\ref{2p}), reproduces the critical behavior of undirected and
directed networks and the phase diagram of percolation
process in complex networks with dependency edges.
Thus, the phase diagram on Fig.~(\ref{f2}) is not a unique
property of the networks with dependency links. A similar
sequence of a second-order transition, the tricritical
point, and a first-order phase transition can be observed
in other physical systems, if their free energy is given by
Eqs.~(\ref{1p}) and (\ref{2p}). Summary of critical behavior near
critical points of continuous and mixed-order phase transitions
in complex networks studied in this paper is presented
in the Table~\ref{tab:table1}.

\section{\label{sec: conclusion} Conclusion}

In this paper, we introduced the susceptibility of arbitrary random undirected and directed complex networks, networks with dependency links, and $k$-cores of networks. The importance of the characteristics is that a strong increase of the susceptibility is the early-warning signal of approaching to the critical point of continuous and mixed-order percolation phase transitions. While finite size effects smooth the transition point of emergence of the giant connected component, they do not strongly impact the maximum of susceptibility. To calculate the characteristics we  introduced so called `observers', which are randomly chosen nodes to monitor the network connectivity and to give information about how many nodes are reachable by following the edges, starting from the observers. Using the data we calculated the susceptibility as the ratio of the change of the number of nodes reachable from the observers to the change of the number of observers. The fraction of the observers serves as the strength of the field conjugated to the giant connected component of a network. We  compared the obtained susceptibility to  the susceptibility found within other approaches and physical models. In the case of undirected networks, at a sufficiently small field, the critical behavior of the susceptibility agrees with the behavior found by use of a ghost field. In contrast to our approach, the ghost field is not a real parameter and must be small enough not to affect percolation. Within our approach, the field strength is a real parameter, which has a meaning of the fraction of the observers. It does not affect percolation across the entire range of values $(0,1]$.      

Usually, when studying percolation in networks, the main focus is on the emergence of the giant connected component, considering finite clusters to be unimportant and insignificant.  Our theoretical analysis demonstrated that the susceptibility is completely determined by the evolution of finite clusters near the percolation point, when  the giant connected component emerges or disappears. To demonstrate efficiency of our method,  we derived explicit equations determining the susceptibility as a function of removed nodes, the strength of the conjugated field, and other control parameters for  uncorrelated random undirected and directed networks and networks with dependency links. 
We considered continuous and mixed-order (first-order) phase transitions in these networks.  The behavior of the susceptibility near the percolation point is determined by the critical properties of the transitions that allows to use the behavior as an indicator of the type of the transitions. Symmetrical or asymmetrical behavior of the susceptibility, a field dependence of the susceptibility maximum and some other peculiarities are fingerprints of the type of the phase transitions and network organization. Finally, we showed that the critical behavior of the susceptibility of complex networks completely agrees with the critical behavior predicted by the phenomenological Landau mean-field theory of phase transitions in application to complex networks. This agreement evidences on the universality of the critical behavior of phase transitions in networks.

It is important to note that our method can be used in numerical simulations of percolation in an arbitrary network. It does not require prior knowledge of the cluster organization of the network and significant computing time. We just use information gathering by the observers distributed randomly over the network. The proposed method  can be applied to study percolation in other kind of networks, for example, correlated networks, networks with a community structure, interdependent networks, networks with higher order interactions, and finite-dimensional networks. 


\begin{acknowledgments}
This work was supported by the i3N associated laboratory LA/P/0037/202, within the scope of the projects UID-B/50025/2020 and UID-P/50025/2020, financed by national funds through the FCT/MEC.
\end{acknowledgments}








\bibliography{susceptibility}

\begin{thebibliography}{26}%
\makeatletter
\providecommand \@ifxundefined [1]{%
 \@ifx{#1\undefined}
}%
\providecommand \@ifnum [1]{%
 \ifnum #1\expandafter \@firstoftwo
 \else \expandafter \@secondoftwo
 \fi
}%
\providecommand \@ifx [1]{%
 \ifx #1\expandafter \@firstoftwo
 \else \expandafter \@secondoftwo
 \fi
}%
\providecommand \natexlab [1]{#1}%
\providecommand \enquote  [1]{``#1''}%
\providecommand \bibnamefont  [1]{#1}%
\providecommand \bibfnamefont [1]{#1}%
\providecommand \citenamefont [1]{#1}%
\providecommand \href@noop [0]{\@secondoftwo}%
\providecommand \href [0]{\begingroup \@sanitize@url \@href}%
\providecommand \@href[1]{\@@startlink{#1}\@@href}%
\providecommand \@@href[1]{\endgroup#1\@@endlink}%
\providecommand \@sanitize@url [0]{\catcode `\\12\catcode `\$12\catcode
  `\&12\catcode `\#12\catcode `\^12\catcode `\_12\catcode `\%12\relax}%
\providecommand \@@startlink[1]{}%
\providecommand \@@endlink[0]{}%
\providecommand \url  [0]{\begingroup\@sanitize@url \@url }%
\providecommand \@url [1]{\endgroup\@href {#1}{\urlprefix }}%
\providecommand \urlprefix  [0]{URL }%
\providecommand \Eprint [0]{\href }%
\providecommand \doibase [0]{http://dx.doi.org/}%
\providecommand \selectlanguage [0]{\@gobble}%
\providecommand \bibinfo  [0]{\@secondoftwo}%
\providecommand \bibfield  [0]{\@secondoftwo}%
\providecommand \translation [1]{[#1]}%
\providecommand \BibitemOpen [0]{}%
\providecommand \bibitemStop [0]{}%
\providecommand \bibitemNoStop [0]{.\EOS\space}%
\providecommand \EOS [0]{\spacefactor3000\relax}%
\providecommand \BibitemShut  [1]{\csname bibitem#1\endcsname}%
\let\auto@bib@innerbib\@empty
\bibitem [{\citenamefont {Stauffer}\ and\ \citenamefont
  {Aharony}(2018)}]{stauffer2003introduction}%
  \BibitemOpen
  \bibfield  {author} {\bibinfo {author} {\bibfnamefont {D.}~\bibnamefont
  {Stauffer}}\ and\ \bibinfo {author} {\bibfnamefont {A.}~\bibnamefont
  {Aharony}},\ }\href {https://doi.org/10.1201/9781315274386} {\emph {\bibinfo
  {title} {Introduction to Percolation Theory}}}\ (\bibinfo  {publisher}
  {Taylor \& Francis},\ \bibinfo {address} {London},\ \bibinfo {year}
  {2018})\BibitemShut {NoStop}%
\bibitem [{\citenamefont {Newman}(2010)}]{newman2010networks}%
  \BibitemOpen
  \bibfield  {author} {\bibinfo {author} {\bibfnamefont {M.}~\bibnamefont
  {Newman}},\ }\href
  {https://doi.org/10.1093/acprof:oso/9780199206650.001.0001} {\emph {\bibinfo
  {title} {Networks: An Introduction}}}\ (\bibinfo  {publisher} {Oxford
  University Press},\ \bibinfo {address} {Oxford},\ \bibinfo {year}
  {2010})\BibitemShut {NoStop}%
\bibitem [{\citenamefont {Dorogovtsev}\ and\ \citenamefont
  {Mendes}(2022)}]{dorogovtsev2022nature}%
  \BibitemOpen
  \bibfield  {author} {\bibinfo {author} {\bibfnamefont {S.~N.}\ \bibnamefont
  {Dorogovtsev}}\ and\ \bibinfo {author} {\bibfnamefont {J.~F.~F.}\
  \bibnamefont {Mendes}},\ }\href
  {https://doi.org/10.1093/oso/9780199695119.001.0001} {\emph {\bibinfo {title}
  {The Nature of Complex Networks}}}\ (\bibinfo  {publisher} {Oxford University
  Press},\ \bibinfo {address} {Oxford},\ \bibinfo {year} {2022})\BibitemShut
  {NoStop}%
\bibitem [{\citenamefont {Baxter}(1982)}]{baxter1982exactly}%
  \BibitemOpen
  \bibfield  {author} {\bibinfo {author} {\bibfnamefont {R.~J.}\ \bibnamefont
  {Baxter}},\ }\href
  {https://physics.anu.edu.au/research/ftp/_files/Exactly.pdf} {\emph {\bibinfo
  {title} {Exactly Solved Models in Statistical Mechanics}}}\ (\bibinfo
  {publisher} {Academic Press},\ \bibinfo {address} {London},\ \bibinfo {year}
  {1982})\BibitemShut {NoStop}%
\bibitem [{\citenamefont {Kasteleyn}\ and\ \citenamefont
  {Fortuin}(1969)}]{kasteleyn1969introduction}%
  \BibitemOpen
  \bibfield  {author} {\bibinfo {author} {\bibfnamefont {P.~W.}\ \bibnamefont
  {Kasteleyn}}\ and\ \bibinfo {author} {\bibfnamefont {C.~M.}\ \bibnamefont
  {Fortuin}},\ }\bibfield  {title} {\enquote {\bibinfo {title} {Phase
  transitions in lattice systems with random local properties},}\ }\href
  {https://ui.adsabs.harvard.edu/abs/1969JPSJS..26...11K/abstract} {\bibfield
  {journal} {\bibinfo  {journal} {J. Phys. Soc. Jpn. Suppl.}\ }\textbf
  {\bibinfo {volume} {26}},\ \bibinfo {pages} {11} (\bibinfo {year}
  {1969})}\BibitemShut {NoStop}%
\bibitem [{\citenamefont {Fortuin}\ and\ \citenamefont
  {Kasteleyn}(1972)}]{fortuin1972random}%
  \BibitemOpen
  \bibfield  {author} {\bibinfo {author} {\bibfnamefont {C.~M.}\ \bibnamefont
  {Fortuin}}\ and\ \bibinfo {author} {\bibfnamefont {P.~W.}\ \bibnamefont
  {Kasteleyn}},\ }\bibfield  {title} {\enquote {\bibinfo {title} {{On the
  random-cluster model: I. Introduction and relation to other models}},}\
  }\href {https://doi.org/10.1016/0031-8914(72)90045-6} {\bibfield  {journal}
  {\bibinfo  {journal} {Physica}\ }\textbf {\bibinfo {volume} {57}},\ \bibinfo
  {pages} {536} (\bibinfo {year} {1972})}\BibitemShut {NoStop}%
\bibitem [{\citenamefont {Wu}(1982)}]{wu1982potts}%
  \BibitemOpen
  \bibfield  {author} {\bibinfo {author} {\bibfnamefont {F.-Y.}\ \bibnamefont
  {Wu}},\ }\bibfield  {title} {\enquote {\bibinfo {title} {{The Potts
  model}},}\ }\href {https://doi.org/10.1103/RevModPhys.54.235} {\bibfield
  {journal} {\bibinfo  {journal} {Rev. Mod. Phys.}\ }\textbf {\bibinfo {volume}
  {54}},\ \bibinfo {pages} {235} (\bibinfo {year} {1982})}\BibitemShut
  {NoStop}%
\bibitem [{\citenamefont {Lee}\ \emph {et~al.}(2004)\citenamefont {Lee},
  \citenamefont {Goh}, \citenamefont {Kahng},\ and\ \citenamefont
  {Kim}}]{lee2004evolution}%
  \BibitemOpen
  \bibfield  {author} {\bibinfo {author} {\bibfnamefont {D.-S.}\ \bibnamefont
  {Lee}}, \bibinfo {author} {\bibfnamefont {K.-I.}\ \bibnamefont {Goh}},
  \bibinfo {author} {\bibfnamefont {B.}~\bibnamefont {Kahng}}, \ and\ \bibinfo
  {author} {\bibfnamefont {D.}~\bibnamefont {Kim}},\ }\bibfield  {title}
  {\enquote {\bibinfo {title} {{Evolution of scale-free random graphs: Potts
  model formulation}},}\ }\href
  {https://doi.org/10.1016/j.nuclphysb.2004.06.029} {\bibfield  {journal}
  {\bibinfo  {journal} {Nucl. Phys. B}\ }\textbf {\bibinfo {volume} {696}},\
  \bibinfo {pages} {351} (\bibinfo {year} {2004})}\BibitemShut {NoStop}%
\bibitem [{\citenamefont {Aharony}(1980)}]{aharony1980universal}%
  \BibitemOpen
  \bibfield  {author} {\bibinfo {author} {\bibfnamefont {A.}~\bibnamefont
  {Aharony}},\ }\bibfield  {title} {\enquote {\bibinfo {title} {Universal
  critical amplitude ratios for percolation},}\ }\href
  {10.1103/PhysRevB.22.400} {\bibfield  {journal} {\bibinfo  {journal} {Phys.
  Rev. B}\ }\textbf {\bibinfo {volume} {22}},\ \bibinfo {pages} {400} (\bibinfo
  {year} {1980})}\BibitemShut {NoStop}%
\bibitem [{\citenamefont {Parshani}\ \emph {et~al.}(2011)\citenamefont
  {Parshani}, \citenamefont {Buldyrev},\ and\ \citenamefont
  {Havlin}}]{parshani2011critical}%
  \BibitemOpen
  \bibfield  {author} {\bibinfo {author} {\bibfnamefont {R.}~\bibnamefont
  {Parshani}}, \bibinfo {author} {\bibfnamefont {S.~V.}\ \bibnamefont
  {Buldyrev}}, \ and\ \bibinfo {author} {\bibfnamefont {S.}~\bibnamefont
  {Havlin}},\ }\bibfield  {title} {\enquote {\bibinfo {title} {Critical effect
  of dependency groups on the function of networks},}\ }\href
  {https://doi.org/10.1073/pnas.100840410} {\bibfield  {journal} {\bibinfo
  {journal} {PNAS}\ }\textbf {\bibinfo {volume} {108}},\ \bibinfo {pages}
  {1007} (\bibinfo {year} {2011})}\BibitemShut {NoStop}%
\bibitem [{\citenamefont {Goltsev}\ \emph {et~al.}(2003)\citenamefont
  {Goltsev}, \citenamefont {Dorogovtsev},\ and\ \citenamefont
  {Mendes}}]{goltsev2003critical}%
  \BibitemOpen
  \bibfield  {author} {\bibinfo {author} {\bibfnamefont {A.~V.}\ \bibnamefont
  {Goltsev}}, \bibinfo {author} {\bibfnamefont {S.~N.}\ \bibnamefont
  {Dorogovtsev}}, \ and\ \bibinfo {author} {\bibfnamefont {J.~F.~F.}\
  \bibnamefont {Mendes}},\ }\bibfield  {title} {\enquote {\bibinfo {title}
  {Critical phenomena in networks},}\ }\href
  {https://doi.org/10.1103/PhysRevE.67.026123} {\bibfield  {journal} {\bibinfo
  {journal} {Phys. Rev. E}\ }\textbf {\bibinfo {volume} {67}},\ \bibinfo
  {pages} {026123} (\bibinfo {year} {2003})}\BibitemShut {NoStop}%
\bibitem [{\citenamefont {Dorogovtsev}\ \emph {et~al.}(2008)\citenamefont
  {Dorogovtsev}, \citenamefont {Goltsev},\ and\ \citenamefont
  {Mendes}}]{dorogovtsev2008critical}%
  \BibitemOpen
  \bibfield  {author} {\bibinfo {author} {\bibfnamefont {S.~N.}\ \bibnamefont
  {Dorogovtsev}}, \bibinfo {author} {\bibfnamefont {A.~V.}\ \bibnamefont
  {Goltsev}}, \ and\ \bibinfo {author} {\bibfnamefont {J.~F.~F.}\ \bibnamefont
  {Mendes}},\ }\bibfield  {title} {\enquote {\bibinfo {title} {Critical
  phenomena in complex networks},}\ }\href
  {https://doi.org/10.1103/RevModPhys.80.1275} {\bibfield  {journal} {\bibinfo
  {journal} {Rev. Mod. Phys.}\ }\textbf {\bibinfo {volume} {80}},\ \bibinfo
  {pages} {1275} (\bibinfo {year} {2008})}\BibitemShut {NoStop}%
\bibitem [{\citenamefont {Newman}\ \emph {et~al.}(2001)\citenamefont {Newman},
  \citenamefont {Strogatz},\ and\ \citenamefont {Watts}}]{newman2001random}%
  \BibitemOpen
  \bibfield  {author} {\bibinfo {author} {\bibfnamefont {M.~E.~J.}\
  \bibnamefont {Newman}}, \bibinfo {author} {\bibfnamefont {S.~H.}\
  \bibnamefont {Strogatz}}, \ and\ \bibinfo {author} {\bibfnamefont {D.~J.}\
  \bibnamefont {Watts}},\ }\bibfield  {title} {\enquote {\bibinfo {title}
  {Random graphs with arbitrary degree distributions and their applications},}\
  }\href {https://doi.org/10.1103/PhysRevE.64.026118} {\bibfield  {journal}
  {\bibinfo  {journal} {Phys. Rev. E}\ }\textbf {\bibinfo {volume} {64}},\
  \bibinfo {pages} {026118} (\bibinfo {year} {2001})}\BibitemShut {NoStop}%
\bibitem [{\citenamefont {Cirigliano}\ \emph {et~al.}(2024)\citenamefont
  {Cirigliano}, \citenamefont {Tim{\'a}r},\ and\ \citenamefont
  {Castellano}}]{cirigliano2024scaling}%
  \BibitemOpen
  \bibfield  {author} {\bibinfo {author} {\bibfnamefont {L.}~\bibnamefont
  {Cirigliano}}, \bibinfo {author} {\bibfnamefont {G.}~\bibnamefont
  {Tim{\'a}r}}, \ and\ \bibinfo {author} {\bibfnamefont {C.}~\bibnamefont
  {Castellano}},\ }\bibfield  {title} {\enquote {\bibinfo {title} {{Scaling and
  universality for percolation in random networks: A unified view}},}\ }\href
  {https://doi.org/10.1103/PhysRevE.110.064303} {\bibfield  {journal} {\bibinfo
   {journal} {Phys. Rev. E}\ }\textbf {\bibinfo {volume} {110}},\ \bibinfo
  {pages} {064303} (\bibinfo {year} {2024})}\BibitemShut {NoStop}%
\bibitem [{\citenamefont {Newman}(2003)}]{newman2003structure}%
  \BibitemOpen
  \bibfield  {author} {\bibinfo {author} {\bibfnamefont {M.~E.~J.}\
  \bibnamefont {Newman}},\ }\bibfield  {title} {\enquote {\bibinfo {title} {The
  structure and function of complex networks},}\ }\href
  {https://doi.org/10.1137/S00361445034248} {\bibfield  {journal} {\bibinfo
  {journal} {SIAM Rev.}\ }\textbf {\bibinfo {volume} {45}},\ \bibinfo {pages}
  {167} (\bibinfo {year} {2003})}\BibitemShut {NoStop}%
\bibitem [{\citenamefont {Broder}\ \emph {et~al.}(2000)\citenamefont {Broder},
  \citenamefont {Kumar}, \citenamefont {Maghoul}, \citenamefont {Raghavan},
  \citenamefont {Rajagopalan}, \citenamefont {Stata}, \citenamefont {Tomkins},\
  and\ \citenamefont {Wiener}}]{broder2000graph}%
  \BibitemOpen
  \bibfield  {author} {\bibinfo {author} {\bibfnamefont {A.}~\bibnamefont
  {Broder}}, \bibinfo {author} {\bibfnamefont {R.}~\bibnamefont {Kumar}},
  \bibinfo {author} {\bibfnamefont {F.}~\bibnamefont {Maghoul}}, \bibinfo
  {author} {\bibfnamefont {P.}~\bibnamefont {Raghavan}}, \bibinfo {author}
  {\bibfnamefont {S.}~\bibnamefont {Rajagopalan}}, \bibinfo {author}
  {\bibfnamefont {R.}~\bibnamefont {Stata}}, \bibinfo {author} {\bibfnamefont
  {A.}~\bibnamefont {Tomkins}}, \ and\ \bibinfo {author} {\bibfnamefont
  {J.}~\bibnamefont {Wiener}},\ }\bibfield  {title} {\enquote {\bibinfo {title}
  {Graph structure in the web},}\ }\href
  {https://doi.org/10.1016/S1389-1286(00)00083-9} {\bibfield  {journal}
  {\bibinfo  {journal} {Comput. Netw.}\ }\textbf {\bibinfo {volume} {33}},\
  \bibinfo {pages} {309} (\bibinfo {year} {2000})}\BibitemShut {NoStop}%
\bibitem [{\citenamefont {Dorogovtsev}\ \emph {et~al.}(2001)\citenamefont
  {Dorogovtsev}, \citenamefont {Mendes},\ and\ \citenamefont
  {Samukhin}}]{dorogovtsev2001giant}%
  \BibitemOpen
  \bibfield  {author} {\bibinfo {author} {\bibfnamefont {S.~N.}\ \bibnamefont
  {Dorogovtsev}}, \bibinfo {author} {\bibfnamefont {J.~F.~F.}\ \bibnamefont
  {Mendes}}, \ and\ \bibinfo {author} {\bibfnamefont {A.~N.}\ \bibnamefont
  {Samukhin}},\ }\bibfield  {title} {\enquote {\bibinfo {title} {Giant strongly
  connected component of directed networks},}\ }\href
  {https://doi.org/10.1103/PhysRevE.64.025101} {\bibfield  {journal} {\bibinfo
  {journal} {Phys. Rev. E}\ }\textbf {\bibinfo {volume} {64}},\ \bibinfo
  {pages} {025101} (\bibinfo {year} {2001})}\BibitemShut {NoStop}%
\bibitem [{\citenamefont {Schwartz}\ \emph {et~al.}(2002)\citenamefont
  {Schwartz}, \citenamefont {Cohen}, \citenamefont {Ben-Avraham}, \citenamefont
  {Barab{\'a}si},\ and\ \citenamefont {Havlin}}]{schwartz2002percolation}%
  \BibitemOpen
  \bibfield  {author} {\bibinfo {author} {\bibfnamefont {N.}~\bibnamefont
  {Schwartz}}, \bibinfo {author} {\bibfnamefont {R.}~\bibnamefont {Cohen}},
  \bibinfo {author} {\bibfnamefont {D.}~\bibnamefont {Ben-Avraham}}, \bibinfo
  {author} {\bibfnamefont {A.-L.}\ \bibnamefont {Barab{\'a}si}}, \ and\
  \bibinfo {author} {\bibfnamefont {S.}~\bibnamefont {Havlin}},\ }\bibfield
  {title} {\enquote {\bibinfo {title} {Percolation in directed scale-free
  networks},}\ }\href {https://doi.org/10.1103/PhysRevE.66.015104} {\bibfield
  {journal} {\bibinfo  {journal} {Phys. Rev. E}\ }\textbf {\bibinfo {volume}
  {66}},\ \bibinfo {pages} {015104} (\bibinfo {year} {2002})}\BibitemShut
  {NoStop}%
\bibitem [{\citenamefont {Bogu{\~n}{\'a}}\ and\ \citenamefont
  {Serrano}(2005)}]{boguna2005generalized}%
  \BibitemOpen
  \bibfield  {author} {\bibinfo {author} {\bibfnamefont {M.}~\bibnamefont
  {Bogu{\~n}{\'a}}}\ and\ \bibinfo {author} {\bibfnamefont {M.~{\'A}.}\
  \bibnamefont {Serrano}},\ }\bibfield  {title} {\enquote {\bibinfo {title}
  {Generalized percolation in random directed networks},}\ }\href
  {https://doi.org/10.1103/PhysRevE.72.016106} {\bibfield  {journal} {\bibinfo
  {journal} {Phys. Rev. E}\ }\textbf {\bibinfo {volume} {72}},\ \bibinfo
  {pages} {016106} (\bibinfo {year} {2005})}\BibitemShut {NoStop}%
\bibitem [{\citenamefont {Tim{\'a}r}\ \emph {et~al.}(2017)\citenamefont
  {Tim{\'a}r}, \citenamefont {Goltsev}, \citenamefont {Dorogovtsev},\ and\
  \citenamefont {Mendes}}]{timar2017mapping}%
  \BibitemOpen
  \bibfield  {author} {\bibinfo {author} {\bibfnamefont {G.}~\bibnamefont
  {Tim{\'a}r}}, \bibinfo {author} {\bibfnamefont {A.~V.}\ \bibnamefont
  {Goltsev}}, \bibinfo {author} {\bibfnamefont {S.~N.}\ \bibnamefont
  {Dorogovtsev}}, \ and\ \bibinfo {author} {\bibfnamefont {J.~F.~F.}\
  \bibnamefont {Mendes}},\ }\bibfield  {title} {\enquote {\bibinfo {title}
  {{Mapping the structure of directed networks: Beyond the bow-tie diagram}},}\
  }\href {https://doi.org/10.1103/PhysRevLett.118.078301} {\bibfield  {journal}
  {\bibinfo  {journal} {Phys. Rev. Lett.}\ }\textbf {\bibinfo {volume} {118}},\
  \bibinfo {pages} {078301} (\bibinfo {year} {2017})}\BibitemShut {NoStop}%
\bibitem [{\citenamefont {Tim{\'a}r}\ \emph {et~al.}(2021)\citenamefont
  {Tim{\'a}r}, \citenamefont {Kov{\'a}cs},\ and\ \citenamefont
  {Mendes}}]{timar2021enhanced}%
  \BibitemOpen
  \bibfield  {author} {\bibinfo {author} {\bibfnamefont {G.}~\bibnamefont
  {Tim{\'a}r}}, \bibinfo {author} {\bibfnamefont {G.}~\bibnamefont
  {Kov{\'a}cs}}, \ and\ \bibinfo {author} {\bibfnamefont {J.~F.~F.}\
  \bibnamefont {Mendes}},\ }\bibfield  {title} {\enquote {\bibinfo {title}
  {Enhanced robustness of single-layer networks with redundant dependencies},}\
  }\href {https://doi.org/10.1103/PhysRevE.103.022321} {\bibfield  {journal}
  {\bibinfo  {journal} {Phys. Rev. E}\ }\textbf {\bibinfo {volume} {103}},\
  \bibinfo {pages} {022321} (\bibinfo {year} {2021})}\BibitemShut {NoStop}%
\bibitem [{\citenamefont {Dorogovtsev}\ \emph {et~al.}(2006)\citenamefont
  {Dorogovtsev}, \citenamefont {Goltsev},\ and\ \citenamefont
  {Mendes}}]{dorogovtsev2006kcore}%
  \BibitemOpen
  \bibfield  {author} {\bibinfo {author} {\bibfnamefont {S.~N.}\ \bibnamefont
  {Dorogovtsev}}, \bibinfo {author} {\bibfnamefont {A.~V.}\ \bibnamefont
  {Goltsev}}, \ and\ \bibinfo {author} {\bibfnamefont {J.~F.~F.}\ \bibnamefont
  {Mendes}},\ }\bibfield  {title} {\enquote {\bibinfo {title} {$k$-core
  organization of complex networks},}\ }\href
  {https://doi.org/10.1103/PhysRevLett.96.040601} {\bibfield  {journal}
  {\bibinfo  {journal} {Phys. Rev. Lett.}\ }\textbf {\bibinfo {volume} {96}},\
  \bibinfo {pages} {040601} (\bibinfo {year} {2006})}\BibitemShut {NoStop}%
\bibitem [{\citenamefont {Goltsev}\ \emph {et~al.}(2006)\citenamefont
  {Goltsev}, \citenamefont {Dorogovtsev},\ and\ \citenamefont
  {Mendes}}]{goltsev2006kbootstrap}%
  \BibitemOpen
  \bibfield  {author} {\bibinfo {author} {\bibfnamefont {A.~V.}\ \bibnamefont
  {Goltsev}}, \bibinfo {author} {\bibfnamefont {S.~N.}\ \bibnamefont
  {Dorogovtsev}}, \ and\ \bibinfo {author} {\bibfnamefont {J.~F.~F.}\
  \bibnamefont {Mendes}},\ }\bibfield  {title} {\enquote {\bibinfo {title}
  {{$k$-core (bootstrap) percolation on complex networks: Critical phenomena
  and nonlocal effects}},}\ }\href {https://doi.org/10.1103/PhysRevE.73.056101}
  {\bibfield  {journal} {\bibinfo  {journal} {Phys. Rev. E}\ }\textbf {\bibinfo
  {volume} {73}},\ \bibinfo {pages} {056101} (\bibinfo {year}
  {2006})}\BibitemShut {NoStop}%
\bibitem [{\citenamefont {Bashan}\ \emph {et~al.}(2011)\citenamefont {Bashan},
  \citenamefont {Parshani},\ and\ \citenamefont
  {Havlin}}]{bashan2011percolation}%
  \BibitemOpen
  \bibfield  {author} {\bibinfo {author} {\bibfnamefont {A.}~\bibnamefont
  {Bashan}}, \bibinfo {author} {\bibfnamefont {R.}~\bibnamefont {Parshani}}, \
  and\ \bibinfo {author} {\bibfnamefont {S.}~\bibnamefont {Havlin}},\
  }\bibfield  {title} {\enquote {\bibinfo {title} {Percolation in networks
  composed of connectivity and dependency links},}\ }\href
  {https://doi.org/10.1103/PhysRevE.83.051127} {\bibfield  {journal} {\bibinfo
  {journal} {Phys. Rev. E.}\ }\textbf {\bibinfo {volume} {83}},\ \bibinfo
  {pages} {051127} (\bibinfo {year} {2011})}\BibitemShut {NoStop}%
\bibitem [{\citenamefont {Bashan}\ and\ \citenamefont
  {Havlin}(2011)}]{bashan2011combined}%
  \BibitemOpen
  \bibfield  {author} {\bibinfo {author} {\bibfnamefont {A.}~\bibnamefont
  {Bashan}}\ and\ \bibinfo {author} {\bibfnamefont {S.}~\bibnamefont
  {Havlin}},\ }\bibfield  {title} {\enquote {\bibinfo {title} {The combined
  effect of connectivity and dependency links on percolation of networks},}\
  }\href {https://doi.org/10.1007/s10955-011-0333-5} {\bibfield  {journal}
  {\bibinfo  {journal} {J. Stat. Phys.}\ }\textbf {\bibinfo {volume} {145}},\
  \bibinfo {pages} {686} (\bibinfo {year} {2011})}\BibitemShut {NoStop}%
\bibitem [{\citenamefont {Lin}\ \emph {et~al.}(2017)\citenamefont {Lin},
  \citenamefont {Kang}, \citenamefont {Wang}, \citenamefont {Zhao},
  \citenamefont {Li},\ and\ \citenamefont {Havlin}}]{lin2017robustness}%
  \BibitemOpen
  \bibfield  {author} {\bibinfo {author} {\bibfnamefont {Y.}~\bibnamefont
  {Lin}}, \bibinfo {author} {\bibfnamefont {R.}~\bibnamefont {Kang}}, \bibinfo
  {author} {\bibfnamefont {Z.}~\bibnamefont {Wang}}, \bibinfo {author}
  {\bibfnamefont {Z.}~\bibnamefont {Zhao}}, \bibinfo {author} {\bibfnamefont
  {D.}~\bibnamefont {Li}}, \ and\ \bibinfo {author} {\bibfnamefont
  {S.}~\bibnamefont {Havlin}},\ }\bibfield  {title} {\enquote {\bibinfo {title}
  {Robustness of networks with dependency topology},}\ }\href
  {https://doi.org/10.1209/0295-5075/118/36002} {\bibfield  {journal} {\bibinfo
   {journal} {Europhys. Lett.}\ }\textbf {\bibinfo {volume} {118}},\ \bibinfo
  {pages} {36002} (\bibinfo {year} {2017})}\BibitemShut {NoStop}%
\end{thebibliography}%

\end{document}